\documentclass[aps,prl,twocolumn,superscriptaddress]{revtex4-2}
\usepackage{amsmath}
\usepackage{amssymb}
\usepackage{graphicx}
\usepackage{hyperref}
\hypersetup{colorlinks=true, linkcolor=blue, citecolor=blue, urlcolor=blue}

\newcommand{\Dnm}[2]{\Delta_{#1#2}}

\begin{document}

\title{Quantum Dissipative Paraelectricity}

\author{A. Cano}
\email{andres.cano@neel.cnrs.fr}
\affiliation{Univ. Grenoble Alpes, CNRS, Grenoble INP, Institut N\'eel,
25 Rue des Martyrs, 38042 Grenoble, France}

\date{\today}

\begin{abstract}
Whether a quantum system with a double-well effective potential undergoes spontaneous symmetry breaking depends not only on the potential landscape but also on the kinetics and the coupling with additional degrees of freedom. 
Here we introduce a quasi-exactly solvable model  to study this problem in the context of ferroelectrics, 
with results that apply to a broad class of quantum phase transitions. 
Exploiting the analytical solutions, we provide a strict definition of the quantum paraelectric regime and identify a distinct quantum ferroelectric regime in which symmetry breaking can be realized without tunneling features. 
We then show that explicit symmetry breaking cannot be inferred from the order-parameter Hamiltonian alone, but requires additional couplings. 
This leads us to identify a regime of
\emph{quantum dissipative paraelectricity}, in which observable symmetry breaking is suppressed during the evolution toward the ground state, even when the double-well structure dominates over zero-point quantum fluctuations.
\end{abstract}

\maketitle

The stabilization of a high-symmetry phase by quantum fluctuations is among the
most striking manifestations of quantum mechanics in condensed-matter systems.
The paradigmatic example is quantum paraelectricity, originally identified in SrTiO$_3$ from the low-temperature departure of the dielectric susceptibility from Curie-Weiss behavior~\cite{Muller1979}. 
More broadly, the presence of phonon soft modes whose classical softening ~\cite{Ginzburg1949,
Cochran1960} is likewise arrested before completion \cite{Yagi2007,meier2022} provide a wider manifestation of the same phenomenon, underscoring the role of quantum effects in determining the stability of certain crystal structures \cite{Mauri2015,Mingo2016}. 
In that case, conventional density-functional-theory (DFT) approaches, in which atomic positions are treated as classical variables,  face fundamental limitations in capturing such effects~\cite{rubio21,spaldin22,kresse23}.

A complete quantum treatment, however, raises a deeper question. When the order parameter itself is a quantum variable, its effective Hamiltonian has definite-parity eigenstates and the equilibrium reduced density matrix is always symmetric, so that no observable symmetry breaking can arise from this Hamiltonian alone. The selection of a broken-symmetry state must therefore involve coupling to additional degrees of freedom, and the outcome depends on the nature of this coupling. Specifically, the system may either lock into one of the symmetry-equivalent wells or remain effectively symmetric reflecting the overall double-well structure --- a possibility we term \emph{quantum dissipative paraelectricity}.

Here we introduce a quasi-exactly solvable model to investigate these questions
with full analytical control. Our main results are: (i) a strict, analytically
derived definition of the quantum paraelectric regime, shown to arise from
zero-point motion rather than tunneling; (ii) the identification of a quantum
ferroelectric regime in which symmetry may be broken without tunneling features,
distinct from the familiar tunneling ferroelectric limit; and (iii) a framework
for the role of environmental coupling, leading to the identification of quantum
dissipative paraelectricity as a phase in which dissipation suppresses observable
symmetry breaking. While we frame the discussion in terms of ferroelectrics,
the analysis applies to any quantum order parameter coupled to an environment.

\paragraph{The model.---}

To describe a ferroelectric instability, we introduce the order parameter $u$
associated with inversion-symmetry breaking. In the simplest case, the Landau
effective potential takes the form $U(u) = \tfrac{a}{2}u^2 + \tfrac{b}{4}u^4$,
with $b > 0$, and the classical transition occurs at $a_c = 0$ when the potential
changes from a single well ($a > 0$) to a double well ($a < 0$) with degenerate
minima at $u_\pm = \pm\sqrt{-a/b}$ \cite{LandauLifshitzStatPhys1,Levanyuk2024}.

\begin{figure*}[t!]
\centering
\includegraphics[width=.55\textwidth]{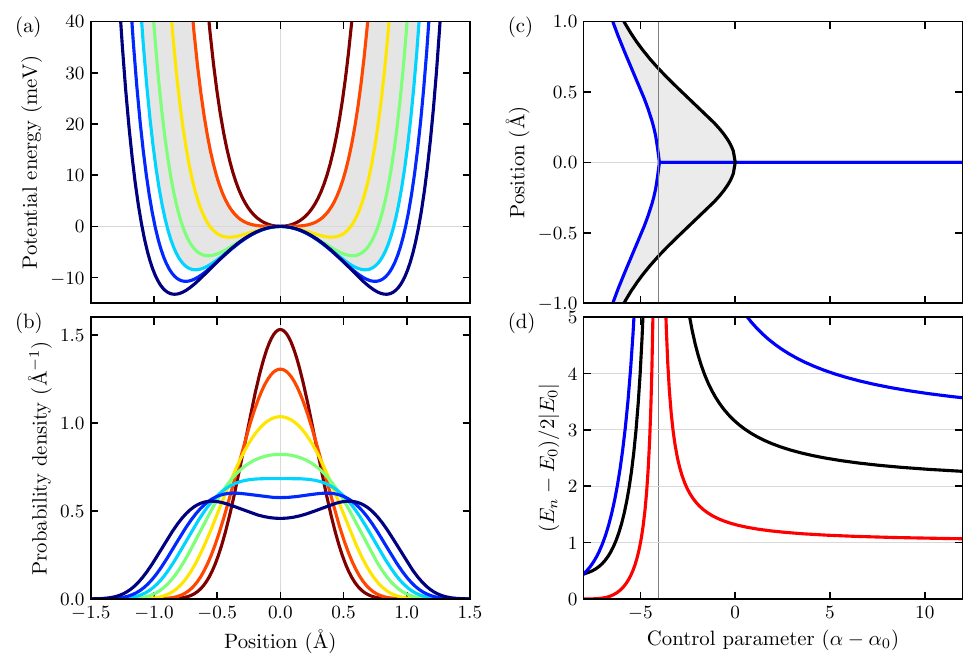}
\caption{%
(a)~Representative shapes of the Razavy potential~\eqref{eq:Razavy} in the
single-well ($\alpha > \alpha_c^{\mathrm{cl}}$) and double-well
($\alpha < \alpha_c^{\mathrm{cl}}$) regimes. The gray region corresponds to the quantum paraelectric regime ($ \alpha_c^{\mathrm{q}} < \alpha < \alpha_c^{\rm cl}$).
(b)~Ground-state probability density $\rho_0(u)$ across the regimes. The
density remains peaked at $u = 0$ throughout the quantum paraelectric phase and
bifurcates only at $\alpha_c^{\mathrm{q}}$, below the classical critical
point $\alpha_c^{\mathrm{cl}}$.
(c)~Locus of the probability-density maximum as a function of $\alpha - \alpha_c^{\rm cl}$,
illustrating the ground-state bifurcation and its separation from the classical
transition (blue vs black curves respectively).
(d)~Relative energies of the three lowest eigenstates vs.\ $\alpha$. The anharmonicity of the potential leads to the nonuniform spacing of the energy levels. Quasi-degenerate doublets appear only well inside the ferroelectric phase
($\alpha \ll \alpha_c^{\mathrm{q}}$), signaling the onset of the tunneling
ferroelectric regime. 
The energy reference is fixed at
$E_0(\alpha_c^{\mathrm{q}}) = 0$.
\label{f:bifurcation}}
\end{figure*}

To incorporate quantum mechanics while retaining full analytical control, we
replace the customary Landau potential with the Razavy
potential~\cite{Razavy1980},
\begin{equation}
U(u) = U_0\alpha\bigl[\alpha - \alpha_c^{\rm cl}
+ \alpha\sinh^2(ku)\bigr]\sinh^2(ku).
\label{eq:Razavy}
\end{equation}
The classical transition is now controlled by $\alpha$, reproducing the Landau form to fourth order in $u$ with $a = 2U_0k^2
\alpha(\alpha-\alpha_c^{\rm cl})$ and $b = \tfrac{4}{3}U_0k^4\alpha(4\alpha-\alpha_c^{\rm cl})$.
For $\alpha > \alpha_c^{\rm cl}$ the potential is a single well, while for $\alpha < \alpha_c^{\rm cl}$ it
becomes a double well with minima at $u_\pm =
\pm\tfrac{1}{2k}\operatorname{arccosh}(\alpha_c^{\rm cl}/\alpha)$ and barrier height
$\Delta U = \tfrac{1}{4}(\alpha-\alpha_c^{\rm cl})^2 U_0$ [see Fig. \ref{f:bifurcation} (a)]. 
We note that the Razavy potential
accurately reproduces DFT potentials calculated for the prototypical materials BaTiO$_3$, SrTiO$_3$, and
KTaO$_3$~\cite{spaldin22}. The fitting parameters are given in the Supplemental
Material.

\paragraph{Quantum phase diagram.---}
The key virtue of the Razavy potential is that the stationary Schr\"odinger
equation,
\begin{equation}
\left(-\frac{\hbar^2}{2m}\frac{d^2}{du^2} + U(u)\right)\psi(u) = E\,\psi(u),
\label{eq:SE}
\end{equation}
admits a finite number of exact closed-form solutions when $U_0 = \hbar^2k^2/(2m)$
and $\alpha_c^{\rm cl} = 2(l+1)$ with $l$ a non-negative
integer~\cite{Razavy1980}, 
providing all the relevant states at low enough temperatures.
Using these solutions, 
the physics of quantum paraelectricity becomes particularly transparent.
For concreteness, 
we consider $l = 3$ 
($\alpha_c^{\rm cl} = 8$). 
Then, the exact
ground-state wave function takes the form $\psi_0(u) =
\phi_0(u)\exp[-\tfrac{\alpha}{4}\cosh(2 ku)]$, with\,\footnote{%
The model has $l+1$ exact eigenstates that can be written as $\psi_n^{(l)}(u) =
\phi_n^{(l)}(u)\exp[-\tfrac{\alpha}{4}\cosh(2 ku)]$, where $\phi_n^{(l)}$ is a
finite polynomial in $\cosh ku$ for even-parity states and $\sinh ku$ for odd ones.}
\begin{equation}
\phi_0(u) = 3\alpha\cosh(ku)
+ \bigl(4-\alpha+2\sqrt{3+(1-\alpha)^2}\bigr)\cosh(3ku),
\label{eq:phi0}
\end{equation}
and the exact ground-state energy is
\begin{equation}
E_0= \frac{\hbar^2 k^2}{2m}\big( 5 - 3\alpha + 2\sqrt{3+(1-\alpha)^2}\big).
\label{eq:E0}
\end{equation}

A central result follows directly from these expressions. 
The bifurcation of the ground-state
probability density $\rho_0(u) = |\psi_0(u)|^2$ does not coincide with
the classical transition at $\alpha_c^{\mathrm{cl}}$ [see Figs.~\ref{f:bifurcation}~(a)-(c)]. 
At
$\alpha_c^{\mathrm{cl}}$, the density remains peaked at $u = 0$ so that
spontaneous symmetry breaking is prevented at the quantum ground-state level. 
Instead,
$\rho_0(u)$ bifurcates only at the lower value
\begin{equation}
\alpha_c^{\mathrm{q}} = \alpha_c^{\mathrm{cl}} - \alpha_{\rm qf},
\label{eq:alphaq}
\end{equation}
where $\alpha_{\rm qf} = (29 - 2\sqrt{19})/5$, corresponding to $E_0 = 0$. This defines the quantum critical
point, and the interval
\begin{equation}
\alpha_c^{\mathrm{q}} < \alpha < \alpha_c^{\mathrm{cl}}
\label{eq:QPEregime}
\end{equation}
provides a strict, analytically grounded definition of the quantum paraelectric
regime. The shift $\alpha_{\rm qf} > 0$ quantifies the stabilization of the
symmetric phase by quantum fluctuations.

\begin{figure*}[!t]
\centering
\includegraphics[width=.95\textwidth]{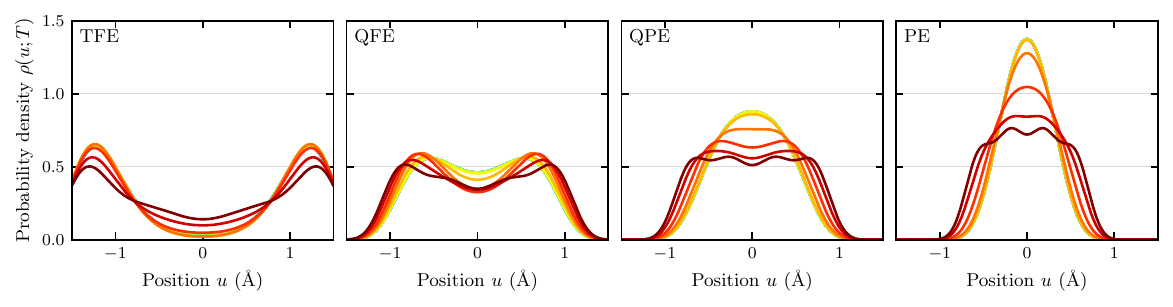}
\caption{%
Thermal probability density $\rho(x;T)$ in the tunneling ferroelectric (TFE),
quantum ferroelectric (QFE), quantum paraelectric (QPE), and standard
paraelectric (PE) regimes. Within each panel, curves correspond to temperatures
logarithmically equally spaced up to
$k_BT = E_3$. In all cases $\rho(x;T)$ remains symmetric, reflecting the
symmetry of the equilibrium density matrix. The central density $\rho(0;T)$
increases with temperature in the TFE panel but decreases in panel QPE, providing
an observable distinction between these two ferroelectric regimes.
\label{f:rhoT}}
\end{figure*}

From this solution, it can also be seen that the shift $\alpha_{\rm qf} $ is a genuine zero-point kinetic-energy effect, rather than a direct consequence of tunneling. The splitting of $\rho_0(u)$ into two maxima requires the wave function to develop additional spatial structure, increasing the kinetic energy through the associated gradients. 
This increase cannot be compensated by the gain in potential energy until the wells become sufficiently deep at $\alpha_c^{\mathrm{q}}$, which defines the domain of existence of quantum paraelectric phase. 
However, this quantum mechanism does not require the system to be in the tunneling regime since the onset of the ferroelectric phase can occur even in the absence of well-defined localized states. This behavior is illustrated in Fig.~\ref{f:bifurcation}~(d).

Another hallmark of tunneling is the presence of quasi-degenerate level pairs with
$\Dnm{0}{1} \equiv E_1 - E_0 \ll E_2 - E_0$, which for the
present model occurs only well inside the ferroelectric phase,
$\alpha \ll \alpha_c^{\mathrm{q}}$, as illustrated in Fig.~\ref{f:bifurcation}~(d). This motivates a classification into two
distinct ferroelectric regimes: a \emph{quantum ferroelectric} regime ($\alpha
\lesssim \alpha_c^{\mathrm{q}}$), in which $\rho_0$ is bifurcated but tunneling
features are absent since the nature of the quantum effects remains the same as in the quantum paralectric phase, and a \emph{tunneling ferroelectric} regime ($\alpha \ll
\alpha_c^{\mathrm{q}}$), in which quasi-degenerate doublets develop and the
tunneling picture applies. These are the quantum counterparts of the displacive
and order-disorder (or Ising) limits of classical phase transitions, respectively.

\paragraph{Symmetry breaking and dissipation.---}

Since the Hamiltonian is parity-symmetric, its eigenstates have definite parity,
and localization in a single well requires a coherent superposition of at least
two states of opposite parity. Such superpositions, however, are not stationary and do not
correspond to equilibrium. In thermal equilibrium, the probability density is
\begin{equation}
\rho(u;T) = \sum_n w_n(T)\,|\psi_n(u)|^2,
\end{equation}
where $w_n = e^{-E_n/k_BT}/\sum_m e^{-E_m/k_BT}$ are Boltzmann weights \cite{CohenTannoudji2019}.
Because each eigenstate has definite parity, $\rho(u;T)$ remains symmetric under
$u \to -u$ at any temperature, even in the presence of quasi-degenerate levels
in the tunneling ferroelectric regime. 
This is illustrated in Fig.~\ref{f:rhoT}.
An useful observable distinction between the tunneling and quantum ferroelectric
regimes emerges from the temperature dependence of $\rho(0;T)$. 
In the quantum ferroelectric regime, the thermal occupation of the antisymmetric first excited state, which is nodal at $u=0$, leads to a relative decrease of $\rho(0;T)$ as temperature increases. 
In contrast, in the tunneling regime the ground and first excited states are quasi-degenerate and therefore remain nearly equally populated over a much broader temperature range. The resulting behavior is then governed by thermal occupation of the symmetric second excited state, which is non-nodal at $u=0$, leading to a relative increase of $\rho(0;T)$ with temperature.
In both cases, however, the two maxima of $\rho(u;T)$ are always equally probable.
Symmetry breaking at this level is therefore only \emph{implicit}---the
probability density bifurcates, but no net polarization develops.

Explicit symmetry breaking~---~a finite observable expectation value $\langle
u \rangle \neq 0$~---~therefore cannot emerge spontaneously from the effective Hamiltonian of the order
parameter alone. It requires a mechanism that breaks the symmetry explicitly,
the most natural being the coupling to additional degrees of freedom acting as an environment.

To see this concretely, consider the system prepared in the paraelectric phase
at temperature $T_0$ and then ``instantaneously'' quenched to zero temperature. The initial
state then can be written as $\Psi(u,t=0) = \sum_n c_n e^{i\theta_n}\psi_n(u)$, where
$c_n = \sqrt{w_n(T_0)}$ and $\theta_n$ are random phases. The time-evolved
density $\rho(u,t) = |\Psi(u,t)|^2$ contains diagonal terms plus off-diagonal interference
terms oscillating at frequencies $(E_n - E_m)/\hbar$. For fully random phases, i.e. thermal equilibrium,
the interference terms vanish upon ensemble averaging and $\langle u \rangle =
0$. The possibility of explicit symmetry breaking is therefore limited to the time interval
before the mixture becomes a fully incoherent.

In the presence of environmental coupling, the off-diagonal elements embodying such a symmetry breaking will survive for some time. 
Retaining the two lowest eigenstates for clarity, which would be a
controlled approximation valid in the $T\to 0$ limit, the probability density after the quenching can be written as
\begin{align}
\rho(u,t)
&= \bigl[c_0^2 + c_1^2(1-e^{-t/\tau_1})\bigr]|\psi_0|^2
+ c_1^2\,|\psi_1|^2 e^{-t/\tau_1} \nonumber\\
&\quad + 2c_0c_1\,\psi_0\psi_1
\cos\!\big(\Dnm{0}{1}t/\hbar\big)e^{-t/\tau_2}.
\label{eq:rho_open}
\end{align}
Here $\tau_1$ denotes the energy-relaxation time while $\tau_2$ is the dephasing time. 
Under standard Markovian weak-coupling assumptions for two-level systems in which there is a clear separation of relaxation and dephasing channels, these times obey the relation 
$
\frac{1}{\tau_2} = \frac{1}{2\tau_1} + \frac{1}{\tau_\phi},
$
where $\tau_\phi$ is a pure dephasing time (such that $\tau_2 = \tau_\phi$ in the pure dephasing limit $\tau_1 \to \infty$ and $\tau_2 \le 2\tau_1$ in general) \cite{Breuer-Petruccione}.
However, in more general situations involving non-Markovian dynamics, correlated noise, strong system-bath coupling, or leakage to higher levels, this separation may break down and the simple additive relation between these times need not apply.
In the ferroelectric context, these times are determined by anharmonic phonon-phonon scattering and phonon-phonon coupling to the
acoustic bath, with $\tau_1$ dominated by inelastic processes and $\tau_2$ by elastic ones.

Whether an observable broken-symmetry state can emerge is determined by the
competition between $\tau_1$, $\tau_2$, and the intrinsic
oscillation time $\tau_{01} \equiv \hbar /\Dnm{0}{1}$.

In the tunneling ferroelectric regime, 
$\tau_{01}$
is naturally long due to the quasi-degeneracy of the low-lying energy levels. 
However, if both dephasing and energy relaxation occur on even longer time scale,
\begin{equation}
 \tau_1^{-1} , \tau_2^{-1} \ll \tau_{01}^{-1}, 
\label{eq:FE_hierarchy}
\end{equation}
the system can become transiently localized in one well, yielding $\langle u
\rangle \neq 0$ over experimentally accessible time scales. 
This scenario can indeed be compatible with the quantum ferroelectric regime, where the larger level splitting implies a shorter intrinsic time scale $\tau_{01}$.
This provides a natural framework for glassy or relaxor-like
ferroelectric responses at low temperature.

Conversely, if dephasing is faster than the intrinsic dynamics,
\begin{equation}
\tau_{01}^{-1} , \tau_1^{-1}   \ll \tau_2^{-1},
\quad \text{ or  }\quad 
\tau_{01}^{-1} \ll \tau_1^{-1} , \tau_2^{-1},
\label{eq:QDPE_hierarchy}
\end{equation}
the overall evolution of the system may prevent to probe any symmetry breaking experimentally. 
In this case, even if the double-well potential dominates over zero-point
fluctuations, only implicit symmetry
breaking can be realized. 
We term this regime \emph{quantum dissipative paraelectricity}, to indicate a paraelectric phase sustained not by quantum fluctuations of the order parameter alone, but also by its coupling to the environment, which in general produces dissipation.

\paragraph{Discussion.---}

\begin{figure}[t!]
\centering
\includegraphics[width=.875\columnwidth]{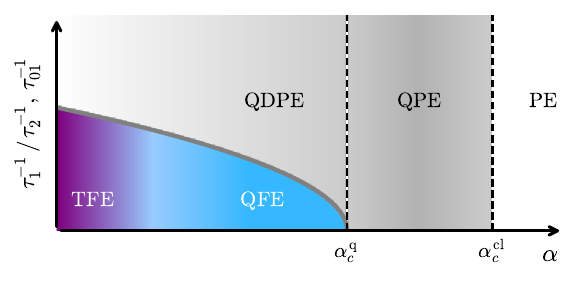}
\caption{%
Schematic phase diagram as a function of the control parameter $\alpha$ and
the environmental coupling strength (parametrized by $\tau_1^{-1}$
relative to $\tau_2^{-1}$ and  $\tau_{01}^{-1}$). 
The standard paraelectric (PE) phase is superseded by the quantum  paraelectric (QPE) phase between $\alpha_c^{\mathrm{cl}}$ and $\alpha_c^{\mathrm{q}}$. 
Below $\alpha_c^{\mathrm{q}}$ in the ferroelectric region, 
the crossover between quantum ferroelectric (QFE) and tunneling ferroelectric (TFE) regimes is determined by the emergence of quasi-degenerate doublets. 
Strong environmental coupling extends the effective paraelectric region by sustaining the quantum dissipative paraelectric (QDPE) phase in the presence of a bifurcated ground-state probability density.
\label{f:phasediagram}}
\end{figure}

The picture that emerges from the above analysis is summarized schematically in
Fig.~\ref{f:phasediagram}. Along the axis of the potential control parameter
$\alpha$, one passes through four regimes as the system is tuned from the
standard paraelectric (PE) phase toward the tunneling ferroelectric phase (TFE). The
quantum paraelectric (QPE) and quantum ferroelectric (QFE) regimes are separated by the
analytically determined critical point $\alpha_c^{\mathrm{q}}$, which lies
strictly below the classical transition $\alpha_c^{\mathrm{cl}}$ by
$\alpha_{\rm qf}$. The occurrence of explicit symmetry
breaking further depends on the coupling to additional degrees of freedom forming an environment, parametrized by the hierarchy of $\tau_1$, $\tau_2$, and $\tau_{01}$.

The quantum dissipative paraelectric (QDPE) phase~\eqref{eq:QDPE_hierarchy} may be
most naturally realized at relatively high
temperatures where thermal fluctuations enhance dissipation or when the coupling to the environment is strong. 
A measurable signature of this regime
would be an anomalously broad or strongly damped soft-mode response: the
dielectric susceptibility would deviate from the quantum paraelectric Barrett
form~\cite{Barrett1952} not through a further decrease of the effective quantum
temperature, but through an increase of the phonon linewidth at low temperature,
reflecting fast relaxation (rather than slow tunneling). 
This overdamped picture has been argued to apply in particular for incommensurate systems, explaining the glass-like thermodynamic properties systematically observed in that case \cite{Cano2004a,Cano2004b,Cano2015}.  
This prediction is in
principle accessible to THz spectroscopy in strained or chemically substituted
variants of SrTiO$_3$ and KTaO$_3$, provided environmental
coupling can be tuned systematically.

The framework also clarifies the status of \emph{ab initio} quantum
calculations of ferroelectrics. In a fully quantum treatment~---~where all
degrees of freedom, including the order parameter, are described by a wave
function~---~the equilibrium density matrix inherits the symmetry of the
Hamiltonian, and the expectation value of $u$ vanishes identically. The
instability of the symmetric state is signaled instead by the bifurcation of
the probability density at $\alpha_c^{\mathrm{q}}$, a feature governed by
zero-point fluctuations and accessible from the exact eigenstates of the model.
Explicit symmetry breaking, by contrast, requires going beyond the effective Hamiltonian
description and specifying the nature of the environmental coupling. The latter
is implicitly present already in classical Landau theory when the potential is
promoted to a free energy with temperature-dependent coefficients~\cite{
Rechester1971,LandauLifshitzStatPhys1,Levanyuk2024}, and becomes explicit and controllable in the present framework.
This fundamental distinction further establishes a bridge between Landau theory and quantum Monte Carlo and path-integral molecular-dynamics simulations.

\paragraph{Conclusions.---}
In summary, we have introduced a quasi-exactly solvable model that allows the
quantum paraelectric regime to be defined sharply and analytically, shown that
this regime is a consequence of zero-point motion rather than tunneling,
identified a quantum ferroelectric regime distinct from the tunneling limit, and
proposed a framework in which the role of dissipation is made explicit. The
resulting quantum dissipative paraelectric phase provides a new mechanism for
the suppression of ferroelectric order beyond the standard quantum fluctuation
scenario.

\paragraph{Acknowledgments.---} I acknowledge extremely useful and insightful discussions with M. Donaire, Q. N. Meier, and A. van Roekeghem.

\bibliography{bib.bib}

\begin{thebibliography}{21}%
\makeatletter
\providecommand \@ifxundefined [1]{%
 \@ifx{#1\undefined}
}%
\providecommand \@ifnum [1]{%
 \ifnum #1\expandafter \@firstoftwo
 \else \expandafter \@secondoftwo
 \fi
}%
\providecommand \@ifx [1]{%
 \ifx #1\expandafter \@firstoftwo
 \else \expandafter \@secondoftwo
 \fi
}%
\providecommand \natexlab [1]{#1}%
\providecommand \enquote  [1]{``#1''}%
\providecommand \bibnamefont  [1]{#1}%
\providecommand \bibfnamefont [1]{#1}%
\providecommand \citenamefont [1]{#1}%
\providecommand \href@noop [0]{\@secondoftwo}%
\providecommand \href [0]{\begingroup \@sanitize@url \@href}%
\providecommand \@href[1]{\@@startlink{#1}\@@href}%
\providecommand \@@href[1]{\endgroup#1\@@endlink}%
\providecommand \@sanitize@url [0]{\catcode `\\12\catcode `\$12\catcode `\&12\catcode `\#12\catcode `\^12\catcode `\_12\catcode `\%12\relax}%
\providecommand \@@startlink[1]{}%
\providecommand \@@endlink[0]{}%
\providecommand \url  [0]{\begingroup\@sanitize@url \@url }%
\providecommand \@url [1]{\endgroup\@href {#1}{\urlprefix }}%
\providecommand \urlprefix  [0]{URL }%
\providecommand \Eprint [0]{\href }%
\providecommand \doibase [0]{https://doi.org/}%
\providecommand \selectlanguage [0]{\@gobble}%
\providecommand \bibinfo  [0]{\@secondoftwo}%
\providecommand \bibfield  [0]{\@secondoftwo}%
\providecommand \translation [1]{[#1]}%
\providecommand \BibitemOpen [0]{}%
\providecommand \bibitemStop [0]{}%
\providecommand \bibitemNoStop [0]{.\EOS\space}%
\providecommand \EOS [0]{\spacefactor3000\relax}%
\providecommand \BibitemShut  [1]{\csname bibitem#1\endcsname}%
\let\auto@bib@innerbib\@empty
\bibitem [{\citenamefont {M{\"u}ller}\ and\ \citenamefont {Burkard}(1979)}]{Muller1979}%
  \BibitemOpen
  \bibfield  {author} {\bibinfo {author} {\bibfnamefont {K.~A.}\ \bibnamefont {M{\"u}ller}}\ and\ \bibinfo {author} {\bibfnamefont {H.}~\bibnamefont {Burkard}},\ }\bibfield  {title} {\bibinfo {title} {{SrTiO$_3$}: An intrinsic quantum paraelectric below 4~{K}},\ }\href {https://doi.org/10.1103/PhysRevB.19.3593} {\bibfield  {journal} {\bibinfo  {journal} {Phys.\ Rev.\ B}\ }\textbf {\bibinfo {volume} {19}},\ \bibinfo {pages} {3593} (\bibinfo {year} {1979})}\BibitemShut {NoStop}%
\bibitem [{\citenamefont {Ginzburg}(1949)}]{Ginzburg1949}%
  \BibitemOpen
  \bibfield  {author} {\bibinfo {author} {\bibfnamefont {V.~L.}\ \bibnamefont {Ginzburg}},\ }\bibfield  {title} {\bibinfo {title} {Theory of ferroelectric phenomena},\ }\href {https://doi.org/https://doi.org/10.3367/UFNr.0038.194908b.0490} {\bibfield  {journal} {\bibinfo  {journal} {Zh. Eksp. Teor. Fiz.}\ }\textbf {\bibinfo {volume} {39}},\ \bibinfo {pages} {490} (\bibinfo {year} {1949})},\ \bibinfo {note} {in Russian}\BibitemShut {NoStop}%
\bibitem [{\citenamefont {Cochran}(1960)}]{Cochran1960}%
  \BibitemOpen
  \bibfield  {author} {\bibinfo {author} {\bibfnamefont {W.}~\bibnamefont {Cochran}},\ }\bibfield  {title} {\bibinfo {title} {Crystal stability and the theory of ferroelectricity},\ }\href {https://doi.org/10.1080/00018736000101229} {\bibfield  {journal} {\bibinfo  {journal} {Adv.\ Phys.}\ }\textbf {\bibinfo {volume} {9}},\ \bibinfo {pages} {387} (\bibinfo {year} {1960})}\BibitemShut {NoStop}%
\bibitem [{\citenamefont {Taniguchi}\ \emph {et~al.}(2007)\citenamefont {Taniguchi}, \citenamefont {Itoh},\ and\ \citenamefont {Yagi}}]{Yagi2007}%
  \BibitemOpen
  \bibfield  {author} {\bibinfo {author} {\bibfnamefont {H.}~\bibnamefont {Taniguchi}}, \bibinfo {author} {\bibfnamefont {M.}~\bibnamefont {Itoh}},\ and\ \bibinfo {author} {\bibfnamefont {T.}~\bibnamefont {Yagi}},\ }\bibfield  {title} {\bibinfo {title} {Ideal soft mode-type quantum phase transition and phase coexistence at quantum critical point in $^{18}\mathrm{O}$-exchanged ${\mathrm{srtio}}_{3}$},\ }\href {https://doi.org/10.1103/PhysRevLett.99.017602} {\bibfield  {journal} {\bibinfo  {journal} {Phys. Rev. Lett.}\ }\textbf {\bibinfo {volume} {99}},\ \bibinfo {pages} {017602} (\bibinfo {year} {2007})}\BibitemShut {NoStop}%
\bibitem [{\citenamefont {Meier}\ \emph {et~al.}()\citenamefont {Meier}, \citenamefont {Mingo},\ and\ \citenamefont {van Roekeghem}}]{meier2022}%
  \BibitemOpen
  \bibfield  {author} {\bibinfo {author} {\bibfnamefont {Q.~N.}\ \bibnamefont {Meier}}, \bibinfo {author} {\bibfnamefont {N.}~\bibnamefont {Mingo}},\ and\ \bibinfo {author} {\bibfnamefont {A.}~\bibnamefont {van Roekeghem}},\ }\href@noop {} {\bibinfo {title} {Finite temperature dielectric properties of ktao$_3$ from first principles and machine learning: Phonon spectra, barrett law, strain engineering and electrostriction}},\ \Eprint {https://arxiv.org/abs/arXiv:2206.08296} {arXiv:2206.08296} \BibitemShut {NoStop}%
\bibitem [{\citenamefont {Errea}\ \emph {et~al.}(2015)\citenamefont {Errea}, \citenamefont {Calandra}, \citenamefont {Pickard}, \citenamefont {Nelson}, \citenamefont {Needs}, \citenamefont {Li}, \citenamefont {Liu}, \citenamefont {Zhang}, \citenamefont {Ma},\ and\ \citenamefont {Mauri}}]{Mauri2015}%
  \BibitemOpen
  \bibfield  {author} {\bibinfo {author} {\bibfnamefont {I.}~\bibnamefont {Errea}}, \bibinfo {author} {\bibfnamefont {M.}~\bibnamefont {Calandra}}, \bibinfo {author} {\bibfnamefont {C.~J.}\ \bibnamefont {Pickard}}, \bibinfo {author} {\bibfnamefont {J.}~\bibnamefont {Nelson}}, \bibinfo {author} {\bibfnamefont {R.~J.}\ \bibnamefont {Needs}}, \bibinfo {author} {\bibfnamefont {Y.}~\bibnamefont {Li}}, \bibinfo {author} {\bibfnamefont {H.}~\bibnamefont {Liu}}, \bibinfo {author} {\bibfnamefont {Y.}~\bibnamefont {Zhang}}, \bibinfo {author} {\bibfnamefont {Y.}~\bibnamefont {Ma}},\ and\ \bibinfo {author} {\bibfnamefont {F.}~\bibnamefont {Mauri}},\ }\bibfield  {title} {\bibinfo {title} {High-pressure hydrogen sulfide from first principles: A strongly anharmonic phonon-mediated superconductor},\ }\href {https://doi.org/10.1103/PhysRevLett.114.157004} {\bibfield  {journal} {\bibinfo  {journal} {Phys. Rev. Lett.}\ }\textbf {\bibinfo {volume} {114}},\ \bibinfo {pages} {157004} (\bibinfo {year} {2015})}\BibitemShut {NoStop}%
\bibitem [{\citenamefont {van Roekeghem}\ \emph {et~al.}(2016)\citenamefont {van Roekeghem}, \citenamefont {Carrete},\ and\ \citenamefont {Mingo}}]{Mingo2016}%
  \BibitemOpen
  \bibfield  {author} {\bibinfo {author} {\bibfnamefont {A.}~\bibnamefont {van Roekeghem}}, \bibinfo {author} {\bibfnamefont {J.}~\bibnamefont {Carrete}},\ and\ \bibinfo {author} {\bibfnamefont {N.}~\bibnamefont {Mingo}},\ }\bibfield  {title} {\bibinfo {title} {Anomalous thermal conductivity and suppression of negative thermal expansion in ${\mathrm{scf}}_{3}$},\ }\href {https://doi.org/10.1103/PhysRevB.94.020303} {\bibfield  {journal} {\bibinfo  {journal} {Phys. Rev. B}\ }\textbf {\bibinfo {volume} {94}},\ \bibinfo {pages} {020303(R)} (\bibinfo {year} {2016})}\BibitemShut {NoStop}%
\bibitem [{\citenamefont {Shin}\ \emph {et~al.}(2021)\citenamefont {Shin}, \citenamefont {Latini}, \citenamefont {Sch\"afer}, \citenamefont {Sato}, \citenamefont {De~Giovannini}, \citenamefont {H\"ubener},\ and\ \citenamefont {Rubio}}]{rubio21}%
  \BibitemOpen
  \bibfield  {author} {\bibinfo {author} {\bibfnamefont {D.}~\bibnamefont {Shin}}, \bibinfo {author} {\bibfnamefont {S.}~\bibnamefont {Latini}}, \bibinfo {author} {\bibfnamefont {C.}~\bibnamefont {Sch\"afer}}, \bibinfo {author} {\bibfnamefont {S.~A.}\ \bibnamefont {Sato}}, \bibinfo {author} {\bibfnamefont {U.}~\bibnamefont {De~Giovannini}}, \bibinfo {author} {\bibfnamefont {H.}~\bibnamefont {H\"ubener}},\ and\ \bibinfo {author} {\bibfnamefont {A.}~\bibnamefont {Rubio}},\ }\bibfield  {title} {\bibinfo {title} {Quantum paraelectric phase of ${\mathrm{srtio}}_{3}$ from first principles},\ }\href {https://doi.org/10.1103/PhysRevB.104.L060103} {\bibfield  {journal} {\bibinfo  {journal} {Phys. Rev. B}\ }\textbf {\bibinfo {volume} {104}},\ \bibinfo {pages} {L060103} (\bibinfo {year} {2021})}\BibitemShut {NoStop}%
\bibitem [{\citenamefont {Esswein}\ and\ \citenamefont {Spaldin}(2022)}]{spaldin22}%
  \BibitemOpen
  \bibfield  {author} {\bibinfo {author} {\bibfnamefont {T.}~\bibnamefont {Esswein}}\ and\ \bibinfo {author} {\bibfnamefont {N.~A.}\ \bibnamefont {Spaldin}},\ }\bibfield  {title} {\bibinfo {title} {Ferroelectric, quantum paraelectric, or paraelectric? calculating the evolution from ${\mathrm{batio}}_{3}$ to ${\mathrm{srtio}}_{3}$ to ${\mathrm{ktao}}_{3}$ using a single-particle quantum mechanical description of the ions},\ }\href {https://doi.org/10.1103/PhysRevResearch.4.033020} {\bibfield  {journal} {\bibinfo  {journal} {Phys. Rev. Res.}\ }\textbf {\bibinfo {volume} {4}},\ \bibinfo {pages} {033020} (\bibinfo {year} {2022})}\BibitemShut {NoStop}%
\bibitem [{\citenamefont {Verdi}\ \emph {et~al.}(2023)\citenamefont {Verdi}, \citenamefont {Ranalli}, \citenamefont {Franchini},\ and\ \citenamefont {Kresse}}]{kresse23}%
  \BibitemOpen
  \bibfield  {author} {\bibinfo {author} {\bibfnamefont {C.}~\bibnamefont {Verdi}}, \bibinfo {author} {\bibfnamefont {L.}~\bibnamefont {Ranalli}}, \bibinfo {author} {\bibfnamefont {C.}~\bibnamefont {Franchini}},\ and\ \bibinfo {author} {\bibfnamefont {G.}~\bibnamefont {Kresse}},\ }\bibfield  {title} {\bibinfo {title} {Quantum paraelectricity and structural phase transitions in strontium titanate beyond density functional theory},\ }\href {https://doi.org/10.1103/PhysRevMaterials.7.L030801} {\bibfield  {journal} {\bibinfo  {journal} {Phys. Rev. Mater.}\ }\textbf {\bibinfo {volume} {7}},\ \bibinfo {pages} {L030801} (\bibinfo {year} {2023})}\BibitemShut {NoStop}%
\bibitem [{\citenamefont {Landau}\ and\ \citenamefont {Lifshitz}(1980)}]{LandauLifshitzStatPhys1}%
  \BibitemOpen
  \bibfield  {author} {\bibinfo {author} {\bibfnamefont {L.~D.}\ \bibnamefont {Landau}}\ and\ \bibinfo {author} {\bibfnamefont {E.~M.}\ \bibnamefont {Lifshitz}},\ }\href@noop {} {\emph {\bibinfo {title} {Statistical Physics, Part 1}}},\ \bibinfo {edition} {3rd}\ ed.,\ \bibinfo {series} {Course of Theoretical Physics}, Vol.~\bibinfo {volume} {5}\ (\bibinfo  {publisher} {Butterworth-Heinemann},\ \bibinfo {address} {Oxford},\ \bibinfo {year} {1980})\BibitemShut {NoStop}%
\bibitem [{\citenamefont {Levanyuk}\ \emph {et~al.}(2024)\citenamefont {Levanyuk}, \citenamefont {Strukov},\ and\ \citenamefont {Cano}}]{Levanyuk2024}%
  \BibitemOpen
  \bibfield  {author} {\bibinfo {author} {\bibfnamefont {A.}~\bibnamefont {Levanyuk}}, \bibinfo {author} {\bibfnamefont {B.}~\bibnamefont {Strukov}},\ and\ \bibinfo {author} {\bibfnamefont {A.}~\bibnamefont {Cano}},\ }\bibfield  {title} {\bibinfo {title} {Ferroelectricity},\ }in\ \href {https://doi.org/https://doi.org/10.1016/B978-0-323-90800-9.00164-5} {\emph {\bibinfo {booktitle} {Encyclopedia of Condensed Matter Physics (Second Edition)}}},\ \bibinfo {editor} {edited by\ \bibinfo {editor} {\bibfnamefont {T.}~\bibnamefont {Chakraborty}}}\ (\bibinfo  {publisher} {Academic Press},\ \bibinfo {address} {Oxford},\ \bibinfo {year} {2024})\ \bibinfo {edition} {second edition}\ ed.,\ pp.\ \bibinfo {pages} {284--296}\BibitemShut {NoStop}%
\bibitem [{\citenamefont {Razavy}(1980)}]{Razavy1980}%
  \BibitemOpen
  \bibfield  {author} {\bibinfo {author} {\bibfnamefont {M.}~\bibnamefont {Razavy}},\ }\bibfield  {title} {\bibinfo {title} {An exactly soluble schrödinger equation with a bistable potential},\ }\href {https://doi.org/10.1119/1.12141} {\bibfield  {journal} {\bibinfo  {journal} {Am. J. Phys.}\ }\textbf {\bibinfo {volume} {48}},\ \bibinfo {pages} {285} (\bibinfo {year} {1980})}\BibitemShut {NoStop}%
\bibitem [{Note1()}]{Note1}%
  \BibitemOpen
  \bibinfo {note} {The model has $l+1$ exact eigenstates that can be written as $\psi _n^{(l)}(u) = \phi _n^{(l)}(u)\exp [-\protect \tfrac {\alpha }{4}\cosh (2 ku)]$, where $\phi _n^{(l)}$ is a finite polynomial in $\cosh ku$ for even-parity states and $\sinh ku$ for odd ones.}\BibitemShut {Stop}%
\bibitem [{\citenamefont {Cohen-Tannoudji}\ \emph {et~al.}(2019)\citenamefont {Cohen-Tannoudji}, \citenamefont {Diu},\ and\ \citenamefont {Lalo{\"e}}}]{CohenTannoudji2019}%
  \BibitemOpen
  \bibfield  {author} {\bibinfo {author} {\bibfnamefont {C.}~\bibnamefont {Cohen-Tannoudji}}, \bibinfo {author} {\bibfnamefont {B.}~\bibnamefont {Diu}},\ and\ \bibinfo {author} {\bibfnamefont {F.}~\bibnamefont {Lalo{\"e}}},\ }\href@noop {} {\emph {\bibinfo {title} {Quantum Mechanics, Volume I: Basic Concepts, Tools, and Applications}}},\ \bibinfo {edition} {2nd}\ ed.\ (\bibinfo  {publisher} {Wiley-VCH},\ \bibinfo {address} {Weinheim},\ \bibinfo {year} {2019})\BibitemShut {NoStop}%
\bibitem [{\citenamefont {Breuer}\ and\ \citenamefont {Petruccione}(2007)}]{Breuer-Petruccione}%
  \BibitemOpen
  \bibfield  {author} {\bibinfo {author} {\bibfnamefont {H.-P.}\ \bibnamefont {Breuer}}\ and\ \bibinfo {author} {\bibfnamefont {F.}~\bibnamefont {Petruccione}},\ }\href {https://doi.org/10.1093/acprof:oso/9780199213900.001.0001} {\emph {\bibinfo {title} {The Theory of Open Quantum Systems}}}\ (\bibinfo  {publisher} {Oxford University Press},\ \bibinfo {year} {2007})\BibitemShut {NoStop}%
\bibitem [{\citenamefont {Barrett}(1952)}]{Barrett1952}%
  \BibitemOpen
  \bibfield  {author} {\bibinfo {author} {\bibfnamefont {J.~H.}\ \bibnamefont {Barrett}},\ }\bibfield  {title} {\bibinfo {title} {Dielectric constant in perovskite type crystals},\ }\href {https://doi.org/10.1103/PhysRev.86.118} {\bibfield  {journal} {\bibinfo  {journal} {Phys.\ Rev.}\ }\textbf {\bibinfo {volume} {86}},\ \bibinfo {pages} {118} (\bibinfo {year} {1952})}\BibitemShut {NoStop}%
\bibitem [{\citenamefont {Cano}\ and\ \citenamefont {Levanyuk}(2004{\natexlab{a}})}]{Cano2004a}%
  \BibitemOpen
  \bibfield  {author} {\bibinfo {author} {\bibfnamefont {A.}~\bibnamefont {Cano}}\ and\ \bibinfo {author} {\bibfnamefont {A.~P.}\ \bibnamefont {Levanyuk}},\ }\bibfield  {title} {\bibinfo {title} {Explanation of the glasslike anomaly in the low-temperature specific heat of incommensurate phases},\ }\href {https://doi.org/10.1103/PhysRevLett.93.245902} {\bibfield  {journal} {\bibinfo  {journal} {Phys. Rev. Lett.}\ }\textbf {\bibinfo {volume} {93}},\ \bibinfo {pages} {245902} (\bibinfo {year} {2004}{\natexlab{a}})}\BibitemShut {NoStop}%
\bibitem [{\citenamefont {Cano}\ and\ \citenamefont {Levanyuk}(2004{\natexlab{b}})}]{Cano2004b}%
  \BibitemOpen
  \bibfield  {author} {\bibinfo {author} {\bibfnamefont {A.}~\bibnamefont {Cano}}\ and\ \bibinfo {author} {\bibfnamefont {A.~P.}\ \bibnamefont {Levanyuk}},\ }\bibfield  {title} {\bibinfo {title} {Low-temperature specific heat of real crystals: Possibility of leading contribution of optical vibrations and short-wavelength acoustical vibrations},\ }\href {https://doi.org/10.1103/PhysRevB.70.212301} {\bibfield  {journal} {\bibinfo  {journal} {Phys. Rev. B}\ }\textbf {\bibinfo {volume} {70}},\ \bibinfo {pages} {212301} (\bibinfo {year} {2004}{\natexlab{b}})}\BibitemShut {NoStop}%
\bibitem [{\citenamefont {Rem\'enyi}\ \emph {et~al.}(2015)\citenamefont {Rem\'enyi}, \citenamefont {Sahling}, \citenamefont {Biljakovi\ifmmode~\acute{c}\else \'{c}\fi{}}, \citenamefont {Stare\ifmmode \check{s}\else \v{s}\fi{}ini\ifmmode~\acute{c}\else \'{c}\fi{}}, \citenamefont {Lasjaunias}, \citenamefont {Lorenzo}, \citenamefont {Monceau},\ and\ \citenamefont {Cano}}]{Cano2015}%
  \BibitemOpen
  \bibfield  {author} {\bibinfo {author} {\bibfnamefont {G.}~\bibnamefont {Rem\'enyi}}, \bibinfo {author} {\bibfnamefont {S.}~\bibnamefont {Sahling}}, \bibinfo {author} {\bibfnamefont {K.}~\bibnamefont {Biljakovi\ifmmode~\acute{c}\else \'{c}\fi{}}}, \bibinfo {author} {\bibfnamefont {D.}~\bibnamefont {Stare\ifmmode \check{s}\else \v{s}\fi{}ini\ifmmode~\acute{c}\else \'{c}\fi{}}}, \bibinfo {author} {\bibfnamefont {J.-C.}\ \bibnamefont {Lasjaunias}}, \bibinfo {author} {\bibfnamefont {J.~E.}\ \bibnamefont {Lorenzo}}, \bibinfo {author} {\bibfnamefont {P.}~\bibnamefont {Monceau}},\ and\ \bibinfo {author} {\bibfnamefont {A.}~\bibnamefont {Cano}},\ }\bibfield  {title} {\bibinfo {title} {Incommensurate systems as model compounds for disorder revealing low-temperature glasslike behavior},\ }\href {https://doi.org/10.1103/PhysRevLett.114.195502} {\bibfield  {journal} {\bibinfo  {journal} {Phys. Rev. Lett.}\ }\textbf {\bibinfo {volume} {114}},\ \bibinfo {pages} {195502} (\bibinfo {year} {2015})}\BibitemShut {NoStop}%
\bibitem [{\citenamefont {Rechester}(1971)}]{Rechester1971}%
  \BibitemOpen
  \bibfield  {author} {\bibinfo {author} {\bibfnamefont {A.~B.}\ \bibnamefont {Rechester}},\ }\bibfield  {title} {\bibinfo {title} {Contribution to the theory of second-order phase transitions at low temperatures},\ }\href@noop {} {\bibfield  {journal} {\bibinfo  {journal} {Sov.\ Phys.\ JETP}\ }\textbf {\bibinfo {volume} {33}},\ \bibinfo {pages} {423} (\bibinfo {year} {1971})},\ \bibinfo {note} {[Zh.\ Eksp.\ Teor.\ Fiz.\ \textbf{60}, 782 (1971)]}\BibitemShut {NoStop}%
\end{thebibliography}%

\newpage 

\
\vspace{3em}

\onecolumngrid
\begin{center}
\textbf{\large Supplemental Material}
\end{center}

\setcounter{equation}{0}
\setcounter{figure}{0}
\setcounter{table}{0}
\renewcommand{\theequation}{S\arabic{equation}}
\renewcommand{\thefigure}{S\arabic{figure}}
\renewcommand{\thetable}{S\arabic{table}}

\
\vspace{-3em}

\subsection*{Fitting parameters for the Razavy potential}

Table~\ref{t:Rparameters} lists the parameters of the Razavy
potential~\eqref{eq:Razavy} that reproduce the first-principles (DFT) effective
potentials for BaTiO$_3$, SrTiO$_3$, and KTaO$_3$ reported in
Ref.~\cite{spaldin22}. All fits use $U_0 = \hbar^2k^2/(2m)$, $\alpha_c^{\rm cl} =
2(l+1)$ with $l=3$ and $m = m_u$ for SrTiO$_3$ and KTaO$_3$ and $m = m_u/2.5$ for BaTiO$_3$. 

\begin{table}[h!]
\centering
\begin{tabular}{lccc}
\hline\hline
 & $\alpha - \alpha_c^{\rm cl}$
 & $k$ (\AA$^{-1}$) \\
\hline
BaTiO$_3$ & $-6.634$ & $10.07$ \\
SrTiO$_3$ & $-3.336$  & $9.934$  \\
KTaO$_3$  & $-0.423$ & $7.737
$ \\
\hline\hline
\end{tabular}
\caption{Razavy-potential parameters reproducing the DFT results of
Ref.~\cite{spaldin22}, for $U_0 = \hbar^2k^2/(2m)$, $\alpha_c^{\rm cl} = 2(l+1)$ with
$l = 3$, and $m = m_u$.
\label{t:Rparameters}}
\end{table}

\end{document}